\documentstyle[seceqn]{elsart}
\begin{document}

\begin{frontmatter}
\title{Finite-Temperature Retarded and Advanced Fields}
\author{H. Arthur Weldon}
\address{Department of Physics, West Virginia University,\\
Morgantown, WV 26506-6315, USA}
\date{\today}

\maketitle

\begin{abstract}
 By employing retarded and advanced propagators, Aurenche and Becherrawy showed how to
rewrite the real-time thermal Feynman rules so that the temperature dependence is removed from
the free propagators and transferred to the vertices. The present paper introduces retarded
and advanced field operators and incorporates all temperature dependence into the
interaction term of the Hamiltonian. Perturbative expansions of the Green functions
in the Hamiltonian formulation give the correct results order by order in perturbation theory.
 The spectrum of the temperature-dependent Hamiltonian  contains the
thermal quasiparticles that produce poles in the propagators. 

\end{abstract}


\end{frontmatter}

\section{ Introduction}

The poles in the free, thermal propagator are at $k^{2}=m^{2}$. However in higher
orders the poles occur at a temperature-dependent effective mass.
This effect was first discovered in self-interacting scalar theories by Dolan and Jackiw [1]
but is now known to be a quite general phenomena. 
  The temperature dependence of the  Higgs boson mass  is responsible
for the phase transition  at large $T$
in the electroweak theory.
 Temperature-dependent masses also occur for particles
with spin.  In  high-temperature QCD  both quarks [2,3] and gluons [3,4] have
effective thermal masses. 
The thermal masses are known to determine  the location of the normal
threshold branch points for particle production [5].

Although  thermal mass effects are ubiquitous in  perturbative computations of Green functions, it is 
not clear how to account for them   in the operator theory since 
 the Hamiltonian contains no temperature.
As an example, for a real scalar field
$\chi$ a typical  Green function that is useful at finite-temperature is the full retarded propagator
\begin{displaymath}
D_{R}^{\prime}(x-y)=-i\theta(x_{0}-y_{0})\sum_{A}{e^{-\beta E_{A}}\over Z_{\beta}}
\langle A|\Big[\chi(x),\chi(y)\Big]|A\rangle.\end{displaymath} 
Unlike the free retarded propagator, the
full retarded propagator 
 depends on $T$ in a very complicated way. The Fourier transform, $D_{R}'(k)$, has poles and cuts in
$k_{0}$ whose  locations are temperature-dependent. How these poles are related to the spectrum of
the Hamiltonian is unclear. The explicit Fourier transform  \begin{displaymath}
D^{\prime}_{R}(k)=\sum_{A,B}(2\pi)^{3}\delta^{3}(\vec{k}+\vec{p}_{A}-\vec{p}_{B})
{|\langle A|\chi(0)|B\rangle|^{2}\over k_{0}+E_{A}-E_{B}+i\eta}
\left({e^{-\beta E_{A}}-e^{-\beta E_{B}}\over
Z_{\beta}}\right)\end{displaymath}
offers no indication that the singularities in $k_{0}$ are temperature-dependent
since $E_{A}$ and $E_{B}$ are the zero-temperature energies of multiparticle states
$|A\rangle$ and $|B\rangle$.

The standard formulations of finite-temperature field 
theory  do not help  explain how the singularities in $k_{0}$ become
temperature-dependent.   The perturbative  Feynman rules
may be derived using either  the path integral [6-8] or operator methods [9,10]
and specify that the vertices 
are the same as at zero temperature and only the propagators are modified by $T$.   
The result for  the  real-time thermal propagator of a
free scalar field is  a $2\times 2$  matrix
\begin{equation}
\Delta_{ab}(k)
=\left(\begin{array}{cc}
(1+n)\Delta_{R} -n\Delta_{A} 
&\hskip0.3cm e^{\sigma k_{0}}n(\Delta_{R}\!-\!\Delta_{A})\\
e^{(\beta-\sigma) k_{0}}n(\Delta_{R}-\Delta_{A})
&\hskip0.3cm  n\Delta_{R}\!-\!(1+n)\Delta_{A}\end{array}\right)_{ab}\end{equation}
in which $\Delta_{R}=1/(k^{2}-m^{2}+ ik_{0}\epsilon)$ and $\Delta_{A}=\Delta_{R}^{*}$ are  the
free retarded and advanced propagators, $n=1/(\exp(\beta k_{0})-1)$ is the Bose-Einstein 
function, and
 $\sigma$ is a free parameter  in the range $0\le\sigma\le\beta$.
The choice  $\sigma=\beta/2$ is common.
Often    (1.1) is
expressed in terms of  the  Feynman propagator
$\Delta_{F}=1/(k^{2}-m^{2}+i\epsilon)$ and its complex conjugate [6-8].
However the retarded and
advanced expressions will be more useful in what follows.

A novel formulation of the Feynman rules was developed by
Aurenche and Becherrawy [11] and will be of central importance here. They showed that the Feynman
rules could be rewritten so that the  propagators are independent of $T$ and the vertices contain all
$T$ dependence. This approach was developed further by van Eijck, Kobes, and van Weert  [12,13].
The notation of [13] will be most convenient here.  The basic idea it to express
(1.1) in the skew-diagonal form 
\begin{equation}
\Delta_{ab}(k)=
V^{T}_{a\beta}(-k_{0})\left(\begin{array}{cc} 0 &\Delta_{A}(k)\\
\Delta_{R}(k) &0\end{array}\right)_{\beta\gamma}V_{\gamma b}(k_{0}).\end{equation}
The important feature is that
  the temperature dependence from the Bose-Einstein function  is contained in the matrices $V(k_{0})$.
This diagonalization leads to a
reformulation of the Feynman rules in which the free propagator is 
the center matrix in  (1.2) containing $\Delta_{R}(k)$ and $\Delta_{A}(k)$.  That 
propagator  matrix is  temperature-independent. The matrices $V(k_{0})$ are incorporated into newly
defined vertices  which are temperature-dependent. 
The transformation automatically gives exactly the same momentum space
integrals as the standard Feynman rules since it is only a rewriting [11-13]. 

The purpose of this paper is to use (1.2) to construct an operator Hamiltonian containing all the
temperature dependence. The new  Hamiltonian should produce exactly the same physical
 effects, and in particular the same n-point functions as
 those obtained by the usual   methods of  finite-temperature field theory. 
The Hamiltonian comes from a Lagrangian and the motivation for the choice of the latter
comes from the path integral for finite-temperature Green functions.  
 Sec 2 explains the choice of the Lagrangian density  and the field variables   $\phi_{R}$
and $\phi_{A}$. The Lagrangian density contains  time derivatives from first order through
N'th order, where N is large but arbitrary.  
 Sec 3 defines the canonical momenta associated with the  N
time derivatives of the fields and constructs the full Hamiltonian operator. As preparation for
what follows, Sec 4 quantizes the free theory with no interaction and  no N'th order time
derivative. Sec 5 quantizes the free theory including the N'th
order time derivative and confirms that the interaction-picture propagators have the correct behavior
in the limit  N $\to\!\infty$.   Sec 6 applies the standard  operator perturbation theory   using
the evolution operator $U(t_{1},t_{2})$ to compute full Green functions as vacuum matrix elements
of time-ordered operator products. In the limit N $\to\!\infty$ the perturbative expansion of
the Green functions is exactly the same as the usual Feynman rules,  thus proving that the
temperature-dependent Hamiltonian describes the same physics. Sec 7 discusses the results and
implications.

\section{Motivation for the  Lagrangian}

This section will explain why the Lagrangian density is chosen to be (2.17) but 
 will not prove that it is correct.  
The proof of (2.17) occurs in  Sec 6, where it is shown that  (2.17) leads to  the correct
perturbative expansion of the Green functions.
To motivate the Lagrangian  it is useful to  start with the path integral
formulation of thermal field theory in real time [6-8]. In real-time formulations it is
necessary to double the number of fields. Instead of one scalar field $\chi$
it is necessary to use two, denoted $\chi_{+}$ and $\chi_{-}$. The  finite-temperature
 generating functional is 
\begin{equation}
Z[j]=\int D\chi_{+}D\chi_{-}\,\exp[i\int d^{4}x ({\cal L}^{c}
+j_{+}\chi_{+}+j_{-}\chi_{-})],\end{equation}
where the contour Lagrangian is
\begin{eqnarray}
{\cal L}^{c}&=&
\half(\partial_{\mu}\chi_{+})(\partial^{\mu}\chi_{+})
-\half m^{2}\chi_{+}^{2}-\lambda(\chi_{+})^{4}\nonumber\\
&-&\half(\partial_{\mu}\chi_{-})(\partial^{\mu}\chi_{-})
+\half m\chi_{-}^{2}
+\lambda(\chi_{-})^{4}.\end{eqnarray}
To absorb the ultraviolet divergences  the parameters $m$ and $\lambda$ should be the
bare mass and bare coupling constant, which should then be written as the physical mass and
coupling plus counter terms.  As ultraviolet
divergences play no  role in what follows, these details  will not be displayed.

\subsection{Choice of $\check{\cal L}_{00}$}

It is straightforward  to express the quartic interactions  as functional derivatives
of the free generating functional [6-8],
$Z_{00}[j]={\rm Det}^{1/2}(i\Delta_{ab})\exp\big(iW[j]\big)$, where $W[j]$ is determined by the
free, thermal propagator $\Delta_{ab}$: \begin{eqnarray}
W[j]&=&-\half\int d^{4}xd^{4}y\,j_{a}(x)
\Delta_{ab}(x-y)j_{b}(y)\nonumber\\
&=&-\half\int {d^{4}k\over(2\pi)^{4}}\,j_{a}(-k)
\Delta_{ab}(k)j_{b}(k)\hskip1cm (a,b=\pm).\end{eqnarray}
This is the usual starting point for deducing the Feynman rules. 

{\it Matrix Transformations:} The notation used will be that of van Eijck, Kobes, and van Weert
 [13]. The free thermal propagator matrix $\Delta_{ab}$ shown in (1.1) has row and columns labeled by
Latin letters $a,b=\pm$ as follows
\begin{equation}
\Delta_{ab}=\left(\begin{array}{cc}\Delta_{++}(k)\hskip0.5cm &\Delta_{+-}(k)\\
\Delta_{-+}(k)\hskip0.5cm & \Delta_{--}(k)\end{array}\right)\end{equation}
(In other references [6-8] the rows
and columns are instead labeled by the integers 1,2.)  The center matrix in (1.2) will be written
\begin{equation}
\Delta_{\alpha\beta}(k)=\left(\begin{array}{cc} \Delta_{RR}(k) &\hskip0.3cm\Delta_{RA}(k)\\
\Delta_{AR}(k) &\Delta_{AA}(k)\end{array}\right)
\equiv\left(\begin{array}{cc} 0&\hskip0.3cm\Delta_{A}(k)\\
\Delta_{R}(k) &0\end{array}\right).\end{equation} 
Thus 
two subscripts are used to specify the  free retarded and advanced propagators $\Delta_{R}(k)$
and $\Delta_{A}(k)$ as matrix elements.
The matrix relation (1.2) is written
  \begin{equation}
\Delta_{ab}(k)=V^{T}_{a\alpha}(-k_{0})\Delta_{\alpha\beta}(k)
V_{\beta b}(k_{0}).\end{equation}
Latin letters $a,b,c,\dots$ always denote $\pm$; Greek letters $\alpha,\beta,\gamma,\dots$ always
denote $R/A$.  The matrix $V_{\beta b}(k_{0})$ has rows labeled by $R/A$ and columns labeled by
$\pm$. The explicit form of $V_{\beta b}(k_{0})$ is contained in [11-13] and is discussed in
Appendix A. As in [13] the  inverse matrix $V^{-1}(k_{0})$ will be denoted by $U^{T}(-k_{0})$ :
\begin{equation}
V(k_{0})U^{T}(-k_{0})=1\hskip2cm U^{T}(-k_{0})V(k_{0})=1.\end{equation}
The rows of $U_{\beta b}(k_{0})$
are labeled by $\beta=R/A$ and columns  by $b=\pm$ just the same as for $V_{\beta b}(k_{0})$.
These automatically satisfy
\begin{equation} U(k_{0})\sigma^{3}U^{T}(-k_{0})=\sigma^{1}
\hskip2cm V(k_{0})\sigma^{3}V^{T}(-k_{0})=\sigma^{1}.\end{equation}

{\it Retarded and Advanced Sources:}
A new version of the generating functional arises if the propagator relation (2.6) is substituted
into (2.3): 
\begin{displaymath}
W[j]=-\half\int{d^{4}k\over (2\pi)^{4}}\;
j_{a}(-k)V^{T}_{a\alpha}(-k_{0})\Delta_{\alpha\beta}(k)V_{\beta b}(k_{0})j_{b}(k).
\end{displaymath}
This suggests that $W$  be considered as a functional of two new  sources
$J_{R/A}$. The transformation from $j_{\pm}$ to $J_{R/A}$ and the inverse transformation are
\begin{equation}
J_{\beta}(k)=V_{\beta b}(k_{0})j_{b}(k)\hskip1cm
J_{\beta}(k)U_{\beta c}(-k_{0})=j_{c}(k)\end{equation}
In terms of the retarded and advanced sources, $W$ is
 \begin{displaymath}
W[J]=-\half\int{d^{4}k\over (2\pi)^{4}}\;
J_{\alpha}(-k)\Delta_{\alpha\beta}(k)J_{\beta}(k).
\end{displaymath}
The two non-vanishing terms in the above integrand are equal since 
 $\Delta_{RA}(k)=\Delta_{AR}(-k)$. The same result in coordinate space is
\begin{displaymath}
W[J]
=-\!\int d^{4}xd^{4}y\,J_{A}(x)\Delta_{R}(x\!-\! y)J_{R}(y).\end{displaymath}

{\it Retarded and Advanced Fields:} The new sources $J_{R/A}$ couple most naturally to 
a new combination of fields. To find this combination, it is only necessary to convert  
the  coupling between the sources and the fields  from coordinate space to
momentum space and then use  (2.9) :
 \begin{eqnarray}
\int\! d^{4}x\, j_{c}(x)\chi_{c}(x)
&=&\int\! {d^{4}k\over (2\pi)^{4}}\,j_{c}(-k)\chi_{c}(k)\nonumber\\
&=&\int\! {d^{4}k\over (2\pi)^{4}}\,J_{\beta}(-k)U_{\beta c}(k_{0})\chi_{c}(k)
.\nonumber\end{eqnarray}
The last line suggests that $U_{\beta c}(k_{0})\chi_{c}(k)$ should be defined as a new field
$\phi_{\beta}$, for $\beta=R$ or $A$. The transformation from $\chi_{\pm}$ to $\phi_{R/A}$ and
the inverse are \begin{equation}
\phi_{\beta}(k)=U_{\beta c}(k_{0})\chi_{c}(k)\hskip1cm
\phi_{\beta}(k)V_{\beta a}(-k_{0})=\chi_{a}(k).\end{equation}
The source term is now expressed as
\begin{displaymath}
\int {d^{4}k\over (2\pi)^{4}}\,J_{\beta}(-k)\phi_{\beta}(k)=
\int
d^{4}x\,\big[J_{R}(x)\phi_{R}(x)+J_{A}(x)\phi_{R}(x)\big]. \end{displaymath}

{\it Free Lagrangian:} The Lagrangian appropriate for the $\phi_{R/A}$ is determined by
examining the free contour action
\begin{displaymath}
I_{00}^{c}=\int d^{4}x\;\big[\half(\partial_{\mu}\chi_{+})(\partial^{\mu}\chi_{+})
-\half(\partial_{\mu}\chi_{-})(\partial^{\mu}\chi_{-})
-\half m^{2}\chi_{+}^{2}+\half m^{2}\chi_{-}^{2}\big].\end{displaymath}
In momentum space the action is
\begin{displaymath}
I_{00}^{c}=\half\int{d^{4}k\over
(2\pi)^{4}}\;(k^{2}-m^{2})\chi_{a}(\!-k)\,\sigma^{3}_{ab}\,\chi_{b}(k). \end{displaymath}
In terms of the new fields $\phi_{R/A}$ the free action is
\begin{displaymath}
I_{00}^{c}=\half\int{d^{4}k\over
(2\pi)^{4}}\;(k^{2}-m^{2})\phi_{\beta}(\!-k)V_{\beta
a}(k_{0})\sigma^{3}_{ab}V^{T}_{b\gamma}(-k_{0})\phi_{\gamma}(k).\end{displaymath}
From the matrix identity (2.8) this simplifies to
\begin{eqnarray}
I_{00}^{c}&=&\half\int {d^{4}k\over (2\pi)^{4}}(k^{2}-m^{2})
\phi_{\beta}(-k)\sigma^{1}_{\beta\gamma}\phi_{\gamma}(k)\nonumber\cr
&=&\int {d^{4}k\over (2\pi)^{4}}(k^{2}-m^{2})
\phi_{R}(-k)\phi_{A}(k).\nonumber
\end{eqnarray}
In coordinate space the action may be written $I_{00}^{c}=\int d^{4}x\,\check{\cal L}_{00}$
where \begin{equation}
\check{\cal L}_{00}
=(\partial_{\mu}\phi_{R})(\partial^{\mu}\phi_{A})
-m^{2}\phi_{R}\phi_{A}.\end{equation}
The reason for the double zero subscript will be explained in Sec 4.
The check ($\;\check{}\;$) appearing in $\check{\cal L}_{00}$ is a reminder that it is the
difference  of two Lagrangian densities. When quantized it will be a Liouville operator [14,15]
that determines the time evolution of quantum amplitudes and of their complex conjugates (as
suggested by the time contour in the path integral).

\subsection{The Full Lagrangian}

The interaction terms in the contour Lagrangian (2.2) involve quartic powers of $\chi_{\pm}$. 
The momentum-space transformation (2.10) from $\chi_{\pm}$ to $\phi_{R/A}$ becomes in coordinate
space
  \begin{equation}
\chi_{\pm}(x)=\int dy_{0}\;\phi_{\alpha}(x_{0}+y_{0},\vec{x})V_{\alpha\pm}(y_{0}),
\end{equation}
where $\alpha$ is summed over $R/A$ and   $V_{\alpha \pm}(y_{0})$ is the Fourier transform of
$V_{\alpha\pm}(k_{0})$. (See Appendix A.)
The interaction terms $(\chi_{\pm})^{4}$  produce
    products of
the field operators $\phi_{R/A}$  at four different times so that ${\cal L}^{c}$ is not
a true Lagrangian density when expressed in terms of $\phi_{R/A}$.
To avoid products of non-simultaneous fields   it is instructive to perform a heuristic
Taylor series expansion.  The  idea is to express 
$\chi_{\pm}$ in terms of the time derivatives of $\phi_{R/A}$ at the same $x_{0}$. 
A formal  expansion of 
 $\phi_{\alpha}(x_{0}+y_{0},\vec{x})$  in (2.12) as a Taylor series in $y_{0}$ gives:
\begin{equation}
\chi_{\pm}(x)\sim  \sum_{\ell=0}^{\infty}
{\partial^{\ell}\phi_{\alpha}(x)\over\partial x_{0}^{\ell}}
C^{\ell}_{\alpha\pm}.\end{equation}
The symbol $\sim$  is a warning that the formal series is  meaningless but it serves to
motivate what follows.  The coefficients $C^{\ell}$ are defined as moments with respect to the
time variable in (2.14)  but they may also be expressed in terms of derivatives in frequency
as in (2.15):
 \begin{eqnarray}
C^{\ell}_{\alpha\pm}&=&\int_{-\infty}^{\infty} dy_{0}\;{(y_{0})^{\ell}
\over\ell!}V_{\alpha\pm}(y_{0})\\
C^{\ell}_{\alpha\pm}&=&{(-i)^{\ell}\over\ell !}\,{\partial^{\ell}V_{\alpha\pm}(k_{0})
\over\partial k_{0}^{\ell}}\Big|_{k_{0}=0}.\end{eqnarray}
The explicit forms of $V_{\alpha\pm}(k_{0})$ shown in Appendix A contain Bose-Einstein functions
of $k_{0}$ and are not infinitely differentiable at $k_{0}=0$. Since the actual value of these
coefficients will not matter until the very end of the calculations in Sec 6, there is no
problem in regularizing the $C^{\ell}$ by introducing a fictitious chemical potential as discussed in
Appendix A. At the end of the calculation the regularization may be removed. In the case of fermion
fields the corresponding $V_{\beta\pm}(k_{0})$ contain Fermi-Dirac functions and are not singular at
$k_{0}=0$.

Using the series (2.13) makes all the terms in $(\chi_{\pm})^{4}$  have the same time.
However the presence of an infinite number of time derivatives is difficult, and probably
impossible, to quantize. The procedure instead will  be to truncate the
derivative expansion at a large order N, construct the Hamiltonian for the truncated problem,
and later take the limit N$\to\infty$. 
 To do this, define the sum of the first N-1 time derivatives as 
\begin{equation}\phi_{\pm}(x)=\sum_{\ell=0}^{N-1}
\;{\partial^{\ell}\phi_{\beta}(x)\over\partial x_{0}^{\ell}}
C^{\ell}_{\beta\pm},\end{equation}
so that if N$\to\infty$ then one expects $\phi_{\pm}$ to approach $\chi_{\pm}$.
The truncation gives a Lagrangian density with quartic interactions $(\phi_{+})^{4}$ and
$(\phi_{-})^{4}$ containing time derivatives up to order N-1. The method for
constructing a Hamiltonian with higher time derivatives will be explained below.
The essential feature is that there will be N-1 canonical momenta. 
To construct the Hamiltonian 
requires inverting the relation between  the highest time derivative and the highest
momentum. This inversion is not manageable since the time derivative of order N-1 is buried in
the series for $\phi_{\pm}$. There is a simple solution: Add an auxiliary 
term containing a higher time derivative of order N. 
The specific Lagrangian density to be used is
\begin{equation}
\check{{\cal L}}=(\partial_{\mu}\phi_{R})(\partial^{\mu}\phi_{A})\!-\!m^{2}\phi_{R}\phi_{A}
-h\,{\partial^{N}\phi_{R}\over\partial x_{0}^{N}}
{\partial^{N}\phi_{A}\over\partial x_{0}^{N}}
\!-\!\lambda(\phi_{+})^{4}\!+\!\lambda(\phi_{-})^{4}.\end{equation}
Now the highest time derivative is N'th order.  Inverting the
relation between the N'th time derivative and the N'th canonical momenta will be trivial. In
order that $h$ has no physical effects it will have to approach zero as N$\to \infty$.
The limiting behavior of $h$ that will be used is
\begin{equation}
h=h_{0}\;N^{-N}.\end{equation}
More general possibilities are discussed following (5.23). 
The field equation that follows from (2.17) is
\begin{equation}
(\partial\!\cdot\!\partial +m^{2})\phi_{\overline{\alpha}}
=-4\lambda\sum_{\ell=0}^{N-1}
(-1)^{\ell}\,C^{\ell}_{\beta a}\sigma^{3}_{ab}
{\partial^{\ell}(\phi_{b})^{3}\over
\partial x_{0}^{\ell}}
-h{\partial^{2N}\phi_{\overline{\alpha}}\over\partial x_{0}^{2N}},
\end{equation}
where $\overline{A}=R$ and $\overline{R}=A$.
The highest time derivative in the field equation is 2N; the highest time derivative of
$\phi_{R/A}$ in the summation over $\ell$  is order 2N-1.

\section{The Full Hamiltonian}

With a finite number of time derivatives in $\check{\cal L}$ it is  straightforward to
construct the Hamiltonian for this system. The necessary generalization was developed by
Ostrogradski in 1850 and is discussed in the mechanics book by E.T. Whittaker [16]. For the
Lagrangian density (2.17) there are canonical momenta 
$\Pi^{1}_{\alpha},\Pi^{2}_{\alpha},\dots,\Pi^{N}_{\alpha}$  and canonical
coordinates
 $\Phi^{1}_{\alpha},\Phi^{2}_{\alpha},\dots,\Phi^{N}_{\alpha}$ where $\alpha=R/A$.
The capital letters $\Pi$ and $\Phi$ will denote Heisenberg operators. (In Sec 4 and
subsequently lower case $\pi$ and $\phi$ will be used for interaction picture operators.)
 The canonical momenta are  
\begin{equation}
\Pi_{\alpha}^{m}=\sum_{\ell=0}^{N-m}(-1)^{\ell}{\partial^{\ell}\over\partial x_{0}^{\ell}}
{\partial \check{\cal L}\over \partial (\partial^{\ell +m} \phi_{\alpha}/\partial 
x_{0}^{\ell+m})}\hskip1cm (m=1,2,\dots,N).\end{equation}
Momenta $\Pi_{\alpha}^{1}, \Pi_{\alpha}^{2},\dots, \Pi_{\alpha}^{N-1}$ all involve the
quartic interaction. Writing these out explicitly is tedious and not very useful.
Only the last momenta has a simple form:
$\Pi_{\alpha}^{N}=-h\,\partial^{N}\phi_{\overline{\alpha}}/\partial x_{0}^{N}$.
The canonical coordinates are the time derivatives of the fields:
\begin{equation}
\Phi_{\alpha}^{\ell}={\partial^{\ell-1}\phi_{\alpha}\over\partial x_{0}^{\ell-1}}
\hskip2cm (\ell=1,2,\dots,N).\end{equation} 
The Hamiltonian density is 
\begin{eqnarray}
\check{{\cal H}}&=&\sum_{\alpha=R/A}\Big(\Pi_{\alpha}^{1}\Phi_{\alpha}^{2}
+\Pi_{\alpha}^{2}\Phi_{\alpha}^{3}+\dots
+\Pi_{\alpha}^{N-1}\Phi_{\alpha}^{N}
+\Pi_{\alpha}^{N}{\partial^{N}\phi_{\alpha}\over\partial x_{0}^{N}}\Big)
\nonumber\\
&&-\check{{\cal L}}(\Phi^{1},\Phi^{2},\dots,\Phi^{N},{\partial^{N}\phi\over\partial
x_{0}^{N}})\end{eqnarray} 
Note that this is  only a Legendre transform with respect to the
highest time derivative. This highest derivative is eliminated using
$\partial^{N}\phi_{\alpha}/\partial x_{0}^{N} =-\Pi^{N}_{\overline{\alpha}}/h$.
The explicit Hamiltonian density is then $\check{{\cal H}}=\check{{\cal H}}_{0}+\check{{\cal
H}}_{I}$ where
\begin{eqnarray}
\check{{\cal H}}_{0}&=&\sum_{\alpha=R/A}\big(\Pi_{\alpha}^{1}\Phi_{\alpha}^{2}
+\Pi_{\alpha}^{2}\Phi_{\alpha}^{3}+\dots
+\Pi_{\alpha}^{N-1}\Phi_{\alpha}^{N}\big)
-{1\over h}\Pi^{N}_{R}\Pi^{N}_{A}\nonumber\\
&&+(\vec{\nabla}\Phi_{R}^{1})\cdot(\vec{\nabla}\Phi_{A}^{1})
+m^{2}\Phi^{1}_{R}\Phi^{1}_{A}-\Phi^{2}_{R}\Phi^{2}_{A}\\
\nonumber\\
\check{{\cal H}}_{I}&=&
\lambda(\Phi_{+})^{4}-\lambda(\Phi_{-})^{4}\end{eqnarray}
It is very important that $\Phi_{\pm}$ are functions only of the canonical coordinates:
\begin{equation}\Phi_{\pm}(x)=\sum_{\ell=0}^{N-1}
\;\Phi_{\beta}^{\ell+1}(x)\;
C^{\ell}_{\beta\pm}.\end{equation}
The nineteenth century development of higher derivative theories  was classical, but if
 the classical Poisson brackets are replaced by commutators the textbook treatment [16]
becomes quantum mechanical. 
The canonical commutation relations are
\begin{equation} \big[\Pi^{m}_{\alpha}(t,\vec{x}), \Phi^{\ell}_{\beta}(t,\vec{y})
\big]={1\over i}\delta^{\ell
m}\,\delta_{\alpha\beta}\,\delta^{3}(\vec{x}-\vec{y}).\end{equation} All other equal-time
commutators among $\Pi$'s and $\Phi$'s vanish.
 The Hamiltonian
$\check{H}=\int d^{3}x\check{{\cal H}}$ generates a complete set of Hamiltonian equations for
the 2N coordinates and 2N momenta: \begin{equation}{\partial \Phi^{\ell}_{\alpha}\over\partial
x_{0}} =i\big[\check{H},\Phi^{\ell}_{\alpha}\big] \hskip1cm
{\partial \Pi^{\ell}_{\alpha}\over\partial x_{0}}
=i\big[\check{H},\Pi^{\ell}_{\alpha}\big]\end{equation}
It is easy to check that this procedure really works.  For each $\alpha$ there are
2N Hamiltonian equations (3.8). Of these, 2N-1 equations express the $\Pi$  and $\Phi$ as time
derivatives of $\Phi^{1}$. The last equation $\dot{\Pi}^{1}=i[\check{H},\Pi^{1}]$
reproduces the   
field equation (2.19).

\section{Prelude: Quantization of $\check{\cal L}_{00}$}

The  construction of field operators that satisfy the Heisenberg equations (3.8) will
be done perturbatively using the interaction picture with free dynamics specified by
$\check{H}_{0}$. The corresponding free Lagrangian density is
\begin{equation}
\check{{\cal L}}_{0}=(\partial_{\mu}\phi_{R})(\partial^{\mu}\phi_{A})\!-\!m^{2}\phi_{R}\phi_{A}
-h\,{\partial^{N}\phi_{R}\over\partial x_{0}^{N}}
{\partial^{N}\phi_{A}\over\partial x_{0}^{N}}.\end{equation}
The presence of the order N time derivatives will be  crucial in  the next sections because 
it enables lower time derivatives of the fields to commute at equal times. The ultimate strategy will
be to compute matrix elements at fixed N and then let N $\to\infty$. To understand why the correct
Feynman diagrams emerge in this limit it is  useful to first quantize the  much simpler
Lagrangian density
 \begin{equation}
\check{{\cal L}}_{00}=(\partial_{\mu}\phi_{R})(\partial^{\mu}\phi_{A})\!-\!m^{2}\phi_{R}\phi_{A}
.\end{equation}
The canonical momenta for this are $\pi_{R}=\partial\phi_{A}/\partial t$ and
$\pi_{A}=\partial\phi_{R}/\partial t$.
The  Hamiltonian density is 
\begin{equation}
\check{\cal H}_{00}=\pi_{A}\pi_{R} +(\vec{\nabla}\phi_{R})\cdot
(\vec{\nabla}\phi_{A})+m^{2}\phi_{R}\phi_{A}.
\end{equation}
The equations of motion that follow from (4.2) are
$(\partial\!\cdot\!\partial +m^{2})\phi_{R}(x)=0$ and
$(\partial\!\cdot\!\partial +m^{2})\phi_{A}(x)=0$.
The solutions to the free field equations can be expanded in terms of spatial Fourier
transforms. For later purposes this will be done in a box of volume $V$ with periodic
boundary conditions that make the wave vectors $\vec{k}$ discrete.  
The free particle energy is $E=(m^{2}+\vec{k}^{2})^{1/2}$:
\begin{eqnarray}
\phi_{A}(t,\vec{x})
={1\over\sqrt{V}}\sum_{\vec{k}}
\big(b^{A}_{\vec{k}}\,e^{i\vec{k}\cdot\vec{x}-iE t}
+c^{A}_{\vec{k}}\,e^{-i\vec{k}\cdot\vec{x}+iE t}\bigr)&\hskip1.5cm &c^{A}=b^{A\dagger}\nonumber\\
 \phi_{R}(t,\vec{x})
={1\over\sqrt{V}}\sum_{\vec{k}}
\big(b^{R}_{\vec{k}}\,e^{i\vec{k}\cdot\vec{x}-iE t}
+c^{R}_{\vec{k}}\,e^{-i\vec{k}\cdot\vec{x}+iE t}\bigr)&
&c^{R}=b^{R\dagger}.
\end{eqnarray}
 Quantization is imposed by the canonical  commutation relations  
\begin{equation}
\Big[\pi_{R}(x),\phi_{R}(y)\Big]_{x_{0}=y_{0}}
=\Big[\pi_{A}(x),\phi_{A}(y)\Big]_{x_{0}=y_{0}}
={1\over i}\delta^{3}(\vec{x}-\vec{x}').\end{equation}
The equal-time commutators of $[\pi_{R},\pi_{A}]$ and $[\phi_{R},\phi_{A}]$ are required to vanish.
These can be satisfied by mode operators that obey
\begin{equation}
\Big[b^{A}_{\vec{k}},c^{R}_{\vec{k}'}\Big]
=\Big[b^{R}_{\vec{k}},c^{A}_{\vec{k}'}\Big]=\delta_{\vec{k},\vec{k}'}/2E.
\end{equation}
All other commutators vanish: $[b_{\vec{k}}^{R},c_{\vec{k}'}^{R}]
=[b_{\vec{k}}^{A},c_{\vec{k}'}^{A}]
=[b_{\vec{k}}^{R},b_{\vec{k}'}^{A}]=0$.
This implies the rather unusual result $[\phi_{R}(x),\phi_{R}(y)]=[\phi_{A}(x),\phi_{A}(y)]=0$ for
any separation $x-y$. The only non-vanishing commutator of the two fields is
\begin{equation}
[\phi_{A}(x),\phi_{R}(y)]=\int {d^{3}k\over 2E(2\pi)^{3}}\,e^{i\vec{k}\cdot(\vec{x}-\vec{y})}
\,\big[e^{-iE(x_{0}-y_{0})}-e^{iE(x_{0}-y_{0})}\big],
\end{equation}
which will be important later.

{\it Retarded and Advanced Vacuum States:} Since the operators $b^{A}$ and $b^{R}$ commute, it
would be conventional to define the vacuum state to be annihilated by these two.  
However it is more useful to define
a vacuum state $|0_{R}\rangle$ that is annihilated by the commuting operators $b^{A}$ and
$c^{A}$:
\begin{equation}
q|0_{R}\rangle=0\hskip1cm {\rm if}\;\;q\in\{ b^{A}, c^{A}\}.\end{equation} 
The state $|0_{R}\rangle$ will be called the retarded vacuum. This state has an infinite norm and
cannot be used to compute expectation values of the form $\langle 0_{R}|{\cal O}|0_{R}\rangle$.
Instead it is necessary to introduce an advanced vacuum $|0_{A}\rangle$.  The advanced vacuum
will be in the bra vector form $\langle 0_{A}|$ to construct matrix elements $\langle
0_{A}|{\cal O}|0_{R}\rangle$ and therefore it is natural to express its definition in
terms of the operators that annihilate the bra state:
\begin{equation}
\langle 0_{A}|q=0\hskip1cm {\rm if}\;\;q\in\{b^{R}, c^{R}\}.\end{equation}
The inner product of the two states can be chosen to be $\langle 0_{A}|0_{R}\rangle=1$.

{\it Free Propagators for $\phi_{R/A}$:}
The free propagators are vacuum matrix elements of
time-ordered products. Since there are two fields $\phi_{R/A}$ the propagator is a $2\times 2$
matrix: \begin{equation}
\Delta_{\alpha\beta}(x-y)=-i\langle 0_{A}|\left(\begin{array}{cc}
T\big\{\phi_{R}(x)\phi_{R}(y)\big\}
&T\big\{\phi_{R}(x)\phi_{A}(y)\big\}\\
T\big\{\phi_{A}(x)\phi_{R}(y)\big\}
&T\big\{\phi_{A}(x)\phi_{A}(y)\big\}
\end{array}\right)|0_{R}\rangle.\end{equation}
All four operator products are time-ordered, just as at zero
temperature. The field operators and the states are all 
temperature-independent.   The evaluation of (4.10) is easy.  The relation $\langle
0_{A}|\phi_{R}=0$ implies that (i) $\Delta_{RR}=0$, (ii) $\Delta_{RA}$ vanishes if
$x_{0}>y_{0}$, and (iii) $\Delta_{AR}$ vanishes if $x_{0}<y_{0}$. The relation
$\phi_{A}|0_{R}\rangle=0$ also implies (ii) and (iii) in addition to (iv) $\Delta_{AA}=0$. The
non-vanishing entries are thus
 \begin{eqnarray}
  \Delta_{AR}(x-y)&=&-i\theta(x_{0}-y_{0})\langle 0_{A}|\phi_{A}(x)\phi_{R}(y)|0_{R}\rangle
\nonumber\\
 \Delta_{RA}(x-y)&=&-i\theta(y_{0}-x_{0})\langle 0_{A}|\phi_{A}(y)\phi_{R}(x)|0_{R}\rangle.
\end{eqnarray}
The matrix elements may  be written as commutators since $\langle
0_{A}|\phi_{A}\phi_{R}|0_{R}\rangle=\langle 0_{A}|\big[\phi_{A},\phi_{R}\big]|0_{R}\rangle$.
The value of the commutator is given in (4.7).
The Fourier representation for the propagator matrix is
 \begin{equation}
\Delta_{\alpha\beta}(x-y)=\int{d^{4}k\over (2\pi)^{4}}\;e^{-ik\cdot (x-y)}
\left(\begin{array}{cc} 0 & \Delta_{RA}(k)\\
\Delta_{AR}(k)& 0\end{array}\right),
\end{equation}
where 
\begin{equation}
\Delta_{AR}(k)=1/(k^{2}-m^{2}+i\epsilon k_{0})\hskip1cm
\Delta_{RA}(k)=1/(k^{2}-m^{2}-i\epsilon k_{0})\end{equation}
are the usual free retarded and advanced propagators.
This is precisely the same as (2.5) and 
 will be the starting point of the perturbation theory developed in the next section.

\section{Interaction Picture: Quantization of $\check{\cal L}_{0}$}

\subsection{Field Operators in the Interaction Picture}

The full Hamiltonian (3.4)-(3.5) is expressed in terms of 
 operators $\Pi$ and $\Phi$
 in the Heisenberg picture. Practical calculations in perturbation theory will require
interaction picture operators $\pi$ and $\phi$, whose time evolution is given by
$\check{H}_{0}$. The corresponding Lagrangian density is
\begin{equation}
\check{{\cal L}}_{0}=(\partial_{\mu}\phi_{R})(\partial^{\mu}\phi_{A})\!-\!m^{2}\phi_{R}\phi_{A}
-h\,{\partial^{N}\phi_{R}\over\partial x_{0}^{N}}
{\partial^{N}\phi_{A}\over\partial x_{0}^{N}}.\end{equation}
Quantization  requires the canonical coordinates
\begin{equation}
\phi^{m}_{\alpha}={\partial^{m-1}\phi_{\alpha}\over\partial t^{m-1}} \hskip3cm
(m=1,2,\dots, N)\end{equation}
and the canonical momenta   
\begin{displaymath}
\pi_{\alpha}^{\ell}=\sum_{s=0}^{N-\ell}(-1)^{s}{\partial^{s}\over\partial x_{0}^{s}}
{\partial \check{\cal L}_{0}\over \partial (\partial^{s +\ell} \phi_{\alpha}/\partial 
x_{0}^{s+\ell})}\hskip1cm (\ell=1,2,\dots,N),\end{displaymath}
where $\alpha=R/A$. In terms of time derivatives, the momenta are
\begin{eqnarray}
\pi^{1}_{\alpha}&=&{\partial\phi_{\overline{\alpha}}\over\partial t}
+(-1)^{N}h{\partial^{2N-1}\phi_{\overline{\alpha}}\over\partial t^{2N-1}}\cr
\pi^{\ell}_{\alpha}&=&(-1)^{N+1-\ell}\,h\,{\partial^{2N-\ell}\phi_{\overline{\alpha}}
\over\partial t^{2N-\ell}}\hskip2cm (\ell=2,3,\dots,N).\end{eqnarray}
As usual the  momenta all commute at equal times, the coordinates commute at equal times, and
the only non-vanishing equal-time commutator is
  \begin{equation}
\big[\pi^{\ell}_{\alpha}(t,\vec{x}),\phi^{m}_{\beta}(t,\vec{y}\big]={1\over i}\delta^{\ell
m} \delta^{3}(\vec{x}-\vec{y}).\end{equation}
This implies that the first 2N-1 time derivatives of $\phi_{A}$ commute at equal times, 
the same for  $\phi_{R}$, and that commutators involving
time derivatives of $\phi_{A}$ and of $\phi_{R}$ must satisfy
 \begin{equation}
\Big[{\partial^{\ell}\phi_{A}(t,\vec{x})\over \partial t^{\ell}}
,{\partial^{m}\phi_{R}(t,\vec{y})\over \partial t^{m}}\Big]
=\cases{ 0 & \hskip-2.2cm if $\ell\!+\!m\neq 2N\!-\!1, 4N\!-\!3,\cdots$\cr\cr
{i\over h}(-1)^{\ell+N}\delta^{3}(\vec{x}-\vec{y})
&if $\ell\!+\!m=2N\!-\!1.$} \end{equation}
The cases in which $\ell+m=4N-3, 6N-5,\dots$ are omitted because they are not independent of  the
case $\ell+m=2N-1$. They are related by   field equation that follows from (5.1):
 \begin{equation}
0=(\partial\!\cdot\!\partial +m^{2})\phi_{\alpha}
+(-1)^{N}h{\partial^{2N}\phi_{\alpha}\over\partial t^{2N}}.
\end{equation}
The field equation is solved by an exponential $\exp(i\vec{k}\cdot\vec{x}-i\omega_{j}t)$ 
 if $\omega_{j}$ satisfies
 \begin{equation}
 0=(\omega_{j})^{2}-E^{2}-h\,(\omega_{j})^{2N}\end{equation}
where $E=(m^{2}+\vec{k}^{2})^{1/2}$. 
This equation has 2N roots. 
 For every root $\omega_{j}$  there  is a root 
$-\omega_{j}$. For $h$ very small, one root is real and close to $E$. This real root will be
called $\omega_{1}$. Thus $\omega_{1}=E+{\cal O}(h)$. Since (5.7) is even, $-\omega_{1}$ is also
a root. For $h$ small, the other $2N-2$ roots are very large: $|\omega_{j}|\approx
h^{-1/(2N-2)}$. For $h\approx N^{-N}$ this means that $|\omega_{j}|\approx N^{1/2}$ for large
$N$. It is convenient to label the roots so that
$\omega_{2},\omega_{3},\dots,\omega_{N}$ are all in the lower half plane. The roots in the
upper half-plane are the negatives of these. The general solution to the field equation may be
written \begin{equation}
\phi_{A}(x)={1\over\sqrt{V}}\sum_{\vec{k}}\sum_{j=1}^{N}\big(b^{A}_{j\vec{k}}\;e^{i\vec{k}\cdot\vec{x}
-i\omega_{j}t}+c^{A}_{j\vec{k}}\;e^{-i\vec{k}\cdot\vec{x}
+i\omega_{j}t}\big).\end{equation}
This form will be used for computing commutators but it disguises the self-adjoint property of
$\phi_{A}$. Since $\omega_{1}$ is real, $c_{1}^{A}=b_{1}^{A\dagger}$.
The other 2N-2 complex roots of (5.7)  come in complex conjugate pairs. Thus for
each one of the  roots $\omega_{2},\omega_{3},\dots,\omega_{N}$ in the lower half-plane  the
complex conjugate  appears on the list
$-\omega_{2},-\omega_{3},\dots,-\omega_{N}$. For example, $\omega_{2}$ is in the lower
half-plane and suppose that $\omega_{2}^{*}=-\omega_{8}$. Then 
$b_{2}^{A}=c_{8}^{A\dagger}$.   This allows $\phi_{A}$ to be self-adjoint even though (5.8)
conceals it. For the retarded field the solution is of the same form:
 \begin{equation}
\phi_{R}(x)={1\over\sqrt{V}}\sum_{\vec{k}}\sum_{j=1}^{N}\big(b^{R}_{j\vec{k}}\;e^{i\vec{k}\cdot\vec{x}
-i\omega_{j}t}
+c^{R}_{j\vec{k}}\;e^{-i\vec{k}\cdot\vec{x}
+i\omega_{j}t}\big).\end{equation}
The equal-time commutators (5.5) can be satisfied as follows.
All the advanced operators, $b^{A}$ and $c^{A}$, must 
commute among themselves. All the retarded operators, $b^{R}$ and $c^{R}$, must
commute among themselves. The only non-vanishing commutators are $[b^{A}_{i},c^{R}_{j}]$ and
$[c^{A}_{i},b^{R}_{j}]$ for $i=j$.  
    The value of this
commutator will depend on the roots of (5.7). The only non-vanishing commutator among the mode
operators is \begin{equation}
\big[b^{A}_{i\vec{k}},c^{R}_{j\vec{k}'}\big]=
\big[b^{R}_{i\vec{k}},c^{A}_{j\vec{k}'}\big]=\delta_{ij}\,Q_{j}
\,\delta_{\vec{k},\vec{k}'},\end{equation}
where $Q_{j}$ is defined as the  residue
of the root $\omega_{j}$ in the lower half-plane:
 \begin{equation} Q_{j}={\omega-\omega_{j}\over
\omega^{2}-E^{2}-h\omega^{2N}}\Bigg|_{\omega=\omega_{j}} ={1\over
2\omega_{j}(1-Nh\omega_{j}^{2N-2})}.\end{equation} 
The $Q_{j}$ satisfy two important identities
\begin{eqnarray}
0&=&\sum_{j=1}^{N}(\omega_{j})^{s}Q_{j}\hskip1cm (s\;{\rm odd}, s\neq 2N\!-\!1, 4N\!-\!3,
6N\!-\!5,\dots)\\
1&=&-2h\sum_{j=1}^{N}(\omega_{j})^{2N-1}Q_{j},\end{eqnarray}
which are proven in Appendix B. These identities implement the commutation relations (5.5) as
follows.
 Using the mode expansions and the commutators (5.10) gives
 \begin{eqnarray}
\Big[{\partial^{\ell}\phi_{A}(t,\vec{x})\over dt^{\ell}},
{\partial^{m}\phi_{R}(t,\vec{y})\over dt^{m}}\Big]=
\Big[(\!-1)^{\ell}\!-\!(\!-1)^{m}\Big]\!\sum_{\vec{k}}{e^{i\vec{k}\cdot
(\vec{x}-\vec{y})}\over V}\sum_{j=1}^{N}
(i\omega_{j})^{\ell+m}Q_{j}.\nonumber\end{eqnarray}
If $\ell$ and $m$ are both even or both odd, this vanishes trivially. If one is even and the other
is odd, then $\ell+m$ is odd and  the sum vanishes because of  (5.12) except when $\ell+m=2N-1,
4N-3,\dots$. For the case $\ell+m=2N-1$ the sum is given by (5.13) and has the correct value
to agree with the canonical requirement (5.5).

\subsection{Free Propagators for $\phi_{R/A}$}

In order for the usual time-ordered perturbation theory to succeed the free propagator
must be the vacuum matrix element of a time-ordered product. As in Sec 4 vacuum matrix elements
will be defined in terms of an advanced bra vector $\langle 0_{A}|$ and a retarded ket vector
$|0_{R}\rangle$. 
The natural manner of defining these  is through the mode operators $b^{R/A}$
and $c^{R/A}$. For a fixed mode operator $q$, either $q|0_{R}\rangle=0$
or $\langle 0_{A}|q=0$. This
 partitions  the mode operators 
 into two disjoint sets. All the operators in each set are mutually commuting. The precise
makeup of these sets is  as follows:  
 \begin{eqnarray}
q\,|0_{R}\rangle&=0\hskip1cm{\rm if}\; q\in\{c_{1}^{A},b_{1}^{A}, b_{2}^{A/R}, b_{3}^{A/R},
\dots,b_{N}^{A/R}\}\cr 
\langle 0_{A}|\,q&=0\hskip1cm{\rm if}\; q\in\{c_{1}^{R},b_{1}^{R}, c_{2}^{A/R}, c_{3}^{A/R},
\dots,c_{N}^{A/R}\}\end{eqnarray}
The roles of $b_{1}$ and of $c_{1}$ are the same as in Sec 4. The remaining operators
are conventional in that  for $j\ge 2$ the $b_{j}$ are destruction operators and the
$c_{j}$ are creation operators. Matrix elements of the form $\langle 0_{A}|q\,q'|0_{R}\rangle$
vanish if $q$ and $q'$ are both $b$'s or both $c$'s. If one is  $b$ and one is  $c$ the matrix
element also vanishes if  both are retarded or both are advanced. The only non-vanishing
vacuum matrix elements of two mode operators are the following:
\begin{eqnarray}
\langle 0_{A}|b_{1\vec{k}}^{A}c_{1\vec{k}'}^{R}|0_{R}\rangle
&=&-\langle 0_{A}|c_{1\vec{k}}^{A}b_{1\vec{k}'}^{R}|0_{R}\rangle
=Q_{1}\delta_{\vec{k}\vec{k}'}\cr
\langle 0_{A}|b_{i\vec{k}}^{A}c_{j\vec{k}'}^{R}|0_{R}\rangle
&=&\;\;\langle 0_{A}|b_{i\vec{k}}^{R}c_{j\vec{k}'}^{A}|0_{R}\rangle
=\delta_{ij}\,Q_{j}\,\delta_{\vec{k}\vec{k}'}\hskip0.5cm  {\rm for}\;\; j=2,3,\dots N.\end{eqnarray}
As a result $\langle 0_{A}|\phi_{A}(x)\phi_{A}(0)|0_{R}\rangle=0$ and 
$\langle 0_{A}|\phi_{R}(x)\phi_{R}(0)|0_{R}\rangle=0$.
The $2\times 2$ propagator matrix is
\begin{equation}
G_{\alpha\beta}(x)=-i\langle 0_{A}|T\left(\phi_{\alpha}(x)\phi_{\beta}(0)\right)
|0_{R}\rangle,\end{equation}
and has vanishing diagonal components:
\begin{equation}
G_{\alpha\beta}(x)=\left(\begin{array}{cc} 0 & G_{RA}(x)\cr
G_{AR}(x) & 0\end{array}\right).\end{equation}
The nonvanishing propagators are  $G_{RA}(x)=-i\langle
0_{A}|T\big(\phi_{R}(x)\phi_{A}(0)\big)|0_{R}\rangle$ and
$G_{AR}(x)=
-i\langle 0_{A}|T\big(\phi_{A}(x)\phi_{R}(0)\big)|0_{R}\rangle$.
By translation invariance $G_{RA}(x)=G_{AR}(-x)$.
The result of evaluating  $G_{AR}(x)$  using (5.15) is 
\begin{eqnarray}
G_{AR}(x)&=&-i\int {d^{3}k\over
(2\pi)^{3}}\,e^{i\vec{k}\cdot\vec{x}}
Q_{1}\theta(t)\,\left(e^{-i\omega_{1}t}-e^{i\omega_{1}t}\right)
\nonumber\\ &&-i\int {d^{3}k\over
(2\pi)^{3}}\,e^{i\vec{k}\cdot\vec{x}}\sum_{j=2}^{N}Q_{j}\left(
\theta(t)\,e^{-i\omega_{j}t}
+\theta(-t)\,e^{i\omega_{j}t}\right).\end{eqnarray}
The time dependence on the second line can be written $\exp(-i\omega_{j}|t|)$.
Since the $\omega_{j}$ for $j=2,3,\dots,N$ have negative imaginary parts  the second falls
exponentially as $t\to\pm\infty$.

{\it Large N Limit:} As $N\to\infty$ the first residue $Q_{1}\to 1/2E$. For $j=2,3,\dots N$ the
behavior is $Q_{j}\to\! -1/(2\omega_{j}N)\sim N^{-3/2}$ since $\omega_{j}\sim N^{1/2}$. 
Thus at large $N$ the $G_{AR}$ propagator becomes the usual retarded free propagator (4.13):
\begin{equation}
\lim_{N\to\infty}G_{AR}(x)=-i\theta(t)
\int {d^{3}k\over 2E(2\pi)^{3}}\;e^{i\vec{k}\cdot\vec{x}}
\left(e^{-iEt}-e^{iEt}\right)=\Delta_{AR}(x)\end{equation}
and similarly for $G_{RA}(x)=G_{AR}(-x)$.

{\it Time Derivatives of the Propagator:} One of the unfamiliar aspects of the theory is that
most time derivatives of the fields  commute at equal times as stated in (5.5).
These vanishing commutators are automatically built into the expression (5.18). 
The first time derivative $\partial G_{AR}(x)/\partial t$ contains no term
proportional to $\delta(t)$. However the second time derivative contains the term
\begin{displaymath} -2\delta(t)\int {d^{3}k\over
(2\pi)^{3}}\;e^{i\vec{k}\cdot\vec{x}}\;\sum_{j=1}^{N}\omega_{j}Q_{j}. \end{displaymath}
This automatically vanishes because of the residue identity (5.12).
One can continue this process and show that $\partial^{\ell}G_{AR}(x)/\partial t^{\ell}$ contains
no $\delta(t)$ term for $\ell<2N$. However for $\ell=2N$ there is a $\delta(t)$ term whose
coefficient is given by the inhomogeneous sum rule (5.13).

\subsection{Free Propagators for $\phi_{\pm}$}

A crucial test of the quantization is  to calculate the temperature-dependent propagators 
\begin{equation}
G_{ab}(x-y)=-i\langle 0_{A}|T\big(\phi_{a}(x)\phi_{b}(y)\big)|
0_{R}\rangle\hskip2cm (a,b=\pm)\end{equation}
since these are the naturally occurring quantities in  perturbative expansions.
The $\phi_{\pm}$ are defined in (3.6) as a sum of the
first $N-1$ time derivatives of $\phi_{R}$ and $\phi_{A}$.
Thus 
\begin{eqnarray}
G_{ab}(x-y)=\sum_{\ell,m=0}^{N-1}
{\partial^{\ell}\over\partial x_{0}^{\ell}}{\partial^{m}\over\partial y_{0}^{m}}
G_{\alpha\beta}(x_{0}\!-\!y_{0},\vec{x}-\vec{y})
C^{\ell}_{\alpha a}C^{m}_{\beta b}.\nonumber
\end{eqnarray} All  time derivatives up to order $N-1$ have been taken outside the time-ordering
because all the equal time commutators  vanish as discussed earlier.
Using the definition (2.13) of the temporal moments $C^{\ell}$ and $C^{m}$ gives
\begin{equation}
G_{ab}(x-y)=\int\! dx_{0}'dy_{0}'\;X_{\alpha\beta}\;V_{\alpha a}(x_{0}')
V_{\beta b}(y_{0}')\end{equation}
\begin{displaymath}
X_{\alpha\beta}\equiv\sum_{\ell,m=0}^{N-1}\left({(x_{0}')^{\ell}\over \ell!}
{\partial^{\ell}\over\partial x_{0}^{\ell}}\right)
\;\left({(y_{0}')^{m}\over m !}
{\partial^{m}\over\partial y_{0}^{m}}\right)
G_{\alpha\beta}(x_{0}\!-\!y_{0},\vec{x}-\vec{y}).\end{displaymath}
In the function $X_{\alpha\beta}$ the $y_{0}$ derivatives can be converted to $x_{0}$
derivatives after which one of the two summations can be performed to give
 \begin{eqnarray}
X_{\alpha\beta}=\sum_{p=0}^{2N-2}{(x_{0}'-y_{0}')^{p}\over p!}
{\partial^{p}\over\partial x_{0}^{p}}\; G_{\alpha\beta}(x_{0}-y_{0},
\vec{x}-\vec{y}).\nonumber\end{eqnarray}
In this form $X_{\alpha\beta}$ is recognizable as  a Taylor series for $G_{\alpha\beta}$ with the
remainder term omitted. Since all the time derivatives of $G_{\alpha\beta}$ up
through order $2N-1$ are continuous (with no $\delta(x_{0}-y_{0})$), Taylor's theorem
gives  a specific form for the remainder $R_{\alpha\beta}$: 
\begin{equation}X_{\alpha\beta}=G_{\alpha\beta}(x_{0}+x_{0}'-y_{0}-y_{0}',\vec{x}-\vec{y})
-R_{\alpha\beta}\end{equation}
\begin{displaymath}
 R_{\alpha\beta}={(x_{0}'-y_{0}')^{2N-1}\over (2N-1)!}\;
{\partial^{2N-1}G_{\alpha\beta}(\xi,\vec{x}-\vec{y})\over
\partial \xi^{2N-1}},\end{displaymath}
where $\xi$ is a time in the range $x_{0}-y_{0}<\xi<x_{0}-y_{0}+x_{0}'-y_{0}'$.
For any value of $N$ there is a $\xi$ such that (5.22) is an exact equality.
By using (5.18) it is easy to see that the remainder vanishes in the large $N$ limit:
\begin{equation}
R_{\alpha\beta}\sim {N(\omega_{j})^{2N-1}Q_{j}\over (2N\!-\!1)!}
\sim {(\omega_{j})^{2N-2}\over (2N\!-\!1)!}\sim {1\over h(2N\!-\!1)!}\to 0.\end{equation}
In the last step the behavior $h\sim N^{-N}$  has been
used. (More general $h$ of the form $h\sim N^{-c N}$ for $c$ in the open interval $0<c<2$ will
also work in that  both the remainder (5.23) and the second line of 
(5.18) will go to zero.) The asymptotic vanishing of the remainder implies that
\begin{eqnarray}\lim_{N\to \infty}X_{\alpha\beta}&=&\lim_{N\to\infty}
G_{\alpha\beta}(x_{0}+x_{0}'-y_{0}-y_{0}',\vec{x}-\vec{y})\nonumber\\
&&\nonumber\\
&=&\Delta_{\alpha\beta}(x_{0}+x_{0}'-y_{0}-y_{0}',\vec{x}-\vec{y}).\nonumber\end{eqnarray}
because of (5.19).
When this is substituted into (5.22)  the time-integration variable can be shifted to a more
convenient form as follows:
\begin{eqnarray}
\lim_{N\to\infty}G_{ab}(x-y)
&=&\int dx_{0}'dy_{0}'\Delta_{\alpha\beta}(x_{0}+x_{0}'-y_{0}-y_{0}',\vec{x}-\vec{y})
V_{\alpha a}(x_{0}')V_{\beta b}(y_{0}')\nonumber\cr
\nonumber\cr
&=&\int\! dx_{0}'dy_{0}'\;\Delta_{\alpha\beta}(x_{0}'-y_{0}',\vec{x}-\vec{y})V_{\alpha
a}(x_{0}'-x_{0})V_{\beta b}(y_{0}'-y_{0}). \nonumber\end{eqnarray}
In momentum space this reads
\begin{equation}
\lim_{N\to\infty}G_{ab}(k)=V_{a\alpha}^{T}(-k_{0})\Delta_{\alpha\beta}(k)V_{\beta
b}(k_{0}).\end{equation}
Comparison to (2.6) and (1.1) shows that the right hand side is indeed the correct free thermal
propagator: \begin{equation}\lim_{N\to\infty}
G_{ab}(k)
=\left(\begin{array}{cc}
(1+n)\Delta_{R} -n\Delta_{A} 
&\hskip0.3cm e^{\sigma k_{0}}n(\Delta_{R}\!-\!\Delta_{A})\\
e^{(\beta-\sigma) k_{0}}n(\Delta_{R}-\Delta_{A})
&\hskip0.3cm  n\Delta_{R}\!-\!(1+n)\Delta_{A}\end{array}\right).\end{equation}

{\it Free Propagators for Mixed Combinations:}

It is sometimes  necessary to evaluate propagators that involve one fundamental field
$\phi_{R/A}$ and one  $\phi_{\pm}$ field.  
The mixed propagator
\begin{equation}
G_{\alpha b}(x-y)=-i\langle 0_{A}|T\big(\phi_{\alpha}(x)\phi_{b}(y)\big)|0_{R}\rangle
\end{equation}
is evaluated in the same way.
The time derivatives can be taken outside the time-ordering. Using the definition of
$C^{\ell}_{\beta b}$ as temporal moments of the thermal matrices $V_{\beta b}$ leads to a
single-variable Taylor series with the result
\begin{displaymath}
\lim_{N\to\infty}G_{\alpha b}(x-y)
= \int dz_{0}\,\Delta_{\alpha\beta}(x_{0}-z_{0},\vec{x}-\vec{y})
V_{\beta b}(z_{0}-y_{0}).
\end{displaymath}
In momentum space this implies
\begin{equation}
\lim_{N\to\infty}G_{\alpha b}(k)=\Delta_{\alpha \beta}(k)V_{\beta b}(k).\end{equation}

\section{Time-Dependent Perturbation Theory}

All the necessary interaction picture propagators have been calculated. It is now straightforward
to apply conventional operator perturbation theory in terms of 
 interaction picture fields. The resulting Green functions will have the property that
when $N\to\infty$ they give the correct perturbative results.
As usual the Heisenberg operators $\Phi_{\alpha}$ are related to the interaction picture operators
$\phi_{\alpha}$ by
\begin{equation}\Phi_{\alpha}(t,\vec{x})=\check{U}(0,t)\phi_{\alpha}(t,\vec{x})\check{U}(t,0),
\end{equation}
where the evolution operator is 
\begin{equation}\check{U}(t,t')=\exp[i\check{H}_{0}t]\exp[-i\check{H}(t-t')]\exp[-i\check{H}_{0}t'].
\end{equation} This may be expanded in terms of operators in the interaction picture:
 \begin{equation}
\check{U}(t,t')=\sum_{n=0}^{\infty}{(-i)^{n}\over n!}\int_{t}^{t'} dt_{1}dt_{2}\dots dt_{n}\,
T\Big[\check{H}_{I}(t_{1})\check{H}_{I}(t_{2})\dots
\check{H}_{I}(t_{n})\Big].\end{equation}
To compute the vacuum matrix elements of time-ordered products
of the interaction picture fields requires a version of Wick's
theorem.

{\it Wick's Theorem:}  To deal with time-ordered products of interaction picture fields it is
convenient to introduce the generating functional
 \begin{equation}
S[j]=\langle 0_{A}|T\exp\big(-\!i\!\int d^{4}x\,j_{\alpha}(x)\phi_{\alpha}(x)\big)
|0_{R}\rangle.\end{equation}
Define $B(x)=j_{\alpha}(x)\phi_{\alpha}(x)$ for brevity. Because the commutator of $B$ with
itself is  a c-number the time-ordered exponential operator may be related to an un-ordered
exponential operator that is multiplied by a c-number: \begin{eqnarray}
T\exp\big(-\!i\!\int d^{4}x\, B(x)\big)&=&\exp\big(-\!i\!\int d^{4}x\, B(x)\big)\cr
&\times&\exp\big(\!-\half\!\int
d^{4}xd^{4}y\,\theta(x_{0}\!-\!y_{0})[B(x),B(y)]\big).\end{eqnarray}
In the plane wave expansions (5.8) and (5.9) of
$\phi_{A}$ and $\phi_{R}$, each of the mode operators  either 
annihilates the retarded vacuum ket, $q|0_{R}\rangle =0$, or annihilates the advanced vacuum 
bra, $\langle 0_{A}|q=0$, as displayed in (5.14). This property provides a unique  decomposition
$B(x)=B_{1}(x)+B_{2}(x)$ where $B_{1}(x)|0_{R}\rangle =0$ and $\langle 0_{A}|B_{2}(x)=0$.
The vacuum matrix element of (6.5) is therefore given by the Baker-Campbell-Hausdorff formula:
\begin{displaymath}
S[j]=\exp\Big(-\half\int d^{4}x\Big(\,[B_{1}(x), B_{2}(y)]
+\theta(x_{0}\!-\!y_{0})[B(x),B(y)]\Big)\Big).\end{displaymath}  
Using $[B_{1}(x),B_{1}(y)]=[B_{2}(x),B_{2}(y)]=0$ this reduces to
\begin{displaymath}
S[j]=\exp\big(\!-\!\int
d^{4}xd^{4}y\,\theta(x_{0}\!-\!y_{0})[B_{1}(x),B_{2}(y)]\big).
\end{displaymath}
The commutator is  
$[B_{1}(x), B_{2}(y)]=\langle 0_{A}|B(x)B(y)|0_{R}\rangle$
so that the exponent can be written as a vacuum matrix element of a time-ordered product
\begin{displaymath}
S[j]=\exp\big(-\half\int d^{4}xd^{4}y\,\langle 0_{A}|T\Big(B(x)B(y)\Big)|0_{R}\rangle
\big).\end{displaymath}
Restoring the definition $B(x)=j_{\alpha}(x)\phi_{\alpha}(x)$  gives the final result
in terms of the propagator (5.16):
\begin{equation}
S[j]=\exp\Big(-\!{1\over 2}\int d^{4}xd^{4}y\,j_{\alpha}(x)G_{\alpha\beta}(x-y)j_{\beta}(y)
\Big).\end{equation}
This is the generating functional for Wick's theorem. 

{\it Sample Commutation:}  A typical computation is the full propagator for the
Heisenberg fields $\Phi_{a}$, $a=\pm$. This is a $2\times 2$ matrix
 \begin{equation}
G'_{ab}(x-y)=-i\langle {\cal O}_{A}|T\big(\Phi_{a}(x)\Phi_{b}(y)\Big)|
{\cal O}_{R}\rangle.\end{equation}
It can be written in terms of interaction picture fields:
\begin{equation}G'_{ab}(x-y)
=-i{\langle 0_{A}|T\big[\phi_{a}(x)\phi_{b}(y)U(\infty,-\infty)\big]|0_{R}\rangle
\over\langle 0_{A}|U(\infty,-\infty)|0_{R}\rangle }.
\end{equation}
This can be expanded in powers of the coupling strength, $\lambda$, as $G'=G^{(0)}
+G^{(1)}+G^{(2)}+\dots$.
The lowest order term, $G^{(0)}$, is the free  propagator computed in Sec 5.3. 
As shown in (5.25)  $G_{ab}^{(0)}$ goes to the usual free propagator
matrix $\Delta_{ab}$ in the limit limit $N\!\to\!\infty$.  The first order correction to 
(6.8) is  the connected part of
 \begin{equation}G^{(1)}_{ab}(x-y)=
\int d^{4}z\langle
0_{A}|T\Big(\phi_{a}(x)[-\lambda\phi_{+}^{4}(z)+\lambda\phi_{-}^{4}(z)]
\phi_{b}(y)\Big)|0_{R}\rangle.\end{equation}  Wick's
theorem (6.6) allows the vacuum matrix element of the time-ordered product of six fields to be
written in terms of vacuum matrix elements of three pairs of fields.  The $\phi_{+}^{4}$ term gives 
\begin{eqnarray} -12\lambda\int d^{4}z\,\langle
0_{A}|T\big(\phi_{a}(x)\phi_{+}(z)\big)|0_{R}\rangle &\langle
0_{A}|T\big(\phi_{+}(z)\phi_{+}(z)\big)|0_{R}\rangle\nonumber\\ \times&\langle
0_{A}|T\big(\phi_{+}(z)\phi_{b}(y)\big)|0_{R}\rangle\nonumber.\end{eqnarray}
After a similar evaluation of the $\phi_{-}^{4}$ term the result is
\begin{eqnarray}
G^{(1)}_{ab}(x-y)=12i\lambda\int d^{4}z
&\Big(&G_{a+}(x-z)G_{++}(z-z)G_{+b}(z-y)\cr
&-&G_{a-}(x-z)G_{--}(z-z)G_{-b}(z-y)\Big)\end{eqnarray}
In momentum space this is
\begin{displaymath}
G^{(1)}_{ab}(k)=12i\lambda\!\int{d^{4}q\over (2\pi)^{4}}
\Big( G_{a+}(k)G_{++}(q)G_{+b}(k)
- G_{a-}(k)G_{--}(q)G_{-b}(k)\Big).\end{displaymath}
Using the result (5.25) the large $N$ limit is expressed in terms of the
usual free propagator $\Delta_{ab}$
\begin{eqnarray}
\lim_{N\to\infty}G^{(1)}_{ab}(k)=12i\lambda \int{d^{4}q\over
(2\pi)^{4}}\Big(&&\Delta_{a+}(k)\Delta_{++}(q)\Delta_{+b}(k)\cr 
-&&\Delta_{a-}(k)\Delta_{--}(q)\Delta_{-b}(k)\Big).\end{eqnarray}
This is  the correct result for the one-loop correction to the thermal
propagator matrix and agrees with the standard answer [6-10].  From this example it is easy to see
that one can compute order by order in the interaction picture and in the limit $N\to\infty$ the
result will coincide with the usual perturbative expansion.

\section{Discussion}

The purpose of constructing the  Hamiltonian formulation of
 finite temperature field theory in terms of $\phi_{R}$ and $\phi_{A}$ 
is  not to obtain  another derivation of the Feynman rules for the n-point functions.
On the contrary, the Feynman rules were developed in order to show that the temperature-dependent
Hamiltonian $\check{H}$ has the same physical content, i.e. exactly the same perturbative
expansion of the Green functions, as the usual formulations.

Any quantum mechanical system allows both wave and particle descriptions. 
Finite-temperature field theory to date has been quite successful with the wave aspects
in that the n-point functions can be calculated perturbatively.
The  path integral formulation places particularly strong emphasis on the wave properties since  the
Feynman rules may be obtained there  without ever introducing the notion of a particle. 
However the particle aspect of the theory cannot be put off indefinitely.
The problem  is that when  particles 
are in a thermal background their self-energies become temperature-dependent.
Consequently a propagator that normally has a pole at $k_{0}^{2}=m^{2}+\vec{k}^{2}$ will instead
have a pole at some $k_{0}$ that is temperature-dependent. Such a pole corresponds to a thermal
quasiparticle. Important evidence that this pole is the proper way to characterize the quasiparticle
comes from the theorem of Kobes, Kunstatter, and Rebhan [17] that in any gauge theory the location
of this pole will be gauge invariant. 

The operator formulation presented above is a preliminary step in understanding the quasiparticle
poles. As discussed in the Introduction, the full finite-temperature retarded propagator
$D_{R}'(k)$ has  a quasiparticle pole in $k_{0}$. As shown in Sec 6 this same propagator may
be  computed as
\begin{equation}
D'_{R}(x-y)=\lim_{N\to\infty}{1\over i}\langle{\cal O}_{A}|T\big(\Phi_{A}(x)\Phi_{R}(y)\big)|
{\cal O}_{R}\rangle.\end{equation}
Because $\Phi$ are Heisenberg fields with their time dependence given by the temperature-dependent
 $\check{H}$, $\Phi(t,\vec{x})=\exp(i\check{H}t)\Phi(0,\vec{x})
\exp(-i\check{H}t)$ and consequently the
 Fourier transform of (7.1) is
\begin{equation}
D'_{R}(k_{0},\vec{x}-\vec{y})=
\lim_{N\to\infty}\langle{\cal O}_{A}|\Phi_{A}(0,\vec{x}){1\over
k_{0}-\check{H}+i\eta}\Phi_{R}(0,\vec{y})| {\cal O}_{R}\rangle.\end{equation}
A future publication will show how the poles in $k_{0}$ come from isolated eigenvalues of
$\check{H}$ and that the branch cuts in $k_{0}$ come from continuum states of two or
more quasiparticles.
Thus $\check{H}$ determines both the time evolution of the
fields and the energy spectrum of the particles in direct analogy with zero temperature.
 It is surprisingly easy 
 to adjust $\check{H}_{0}$ so
that its spectrum coincides with that of the full  Hamiltonian $\check{H}$. 
The shifted $\check{H}_{0}$ will describe free quasiparticles. This is parallel to the
zero-temperature shift that make the free Hamiltonian contain the physical mass rather than the bare
mass. At finite temperature this leads to  a quasiparticle resummation program.
Although the above discussion only treats spinless fields, a future publication will  extend the
method to the more physically interesting case of  fermions and gauge bosons. The ultimate
purpose is to obtain probabilities for quasiparticle processes that are gauge-invariant
and infrared finite.

The methods used in Sec 3 may be useful in other problems. It often happens that effective actions
contains products of fields a different times. The Braaten-Pisarski effective action [18] is
such an example. Sec 3 and the later development shows how to expand the action in a 
finite number N of time derivatives, apply canonical methods to construct the Hamiltonian or
the energy-momentum tensor, more generally, and
quantize the result.

\begin{ack}

  This work was supported in part by the U.S. National
Science Foundation under grant PHY-9630149.

\end{ack}

\appendix

\section{Diagonalization of the Propagator}

The matrix $V_{\alpha b}(k_{0})$ that diagonalizes the free thermal propagator in (1.1) was shown by 
van Eijck, Kobes, and van Weert [13] to be
\begin{equation}
\left(\begin{array}{cc}V_{R+}(k_{0})\em & V_{R-}(k_{0})\\
V_{A+}(k_{0}) & V_{A-}(k_{0})\end{array}\right)
=\left(\begin{array}{cc} e^{\beta k_{0}}n(k_{0})/a(-k_{0})\hskip0.5cm &e^{\sigma k_{0}}
n(k_{0})/a(-k_{0})\\ a(k_{0}) & e^{\sigma k_{0}}a(k_{0})\end{array}\right),\end{equation}
where $a(k_{0})$ is arbitrary. The  matrix $U_{\alpha b}(k_{0})$
that occurs in (2.7) is 
\begin{equation}
\left(\begin{array}{cc}U_{R+}(k_{0})\em & U_{R-}(k_{0})\\
U_{A+}(k_{0}) & U_{A-}(k_{0})\end{array}\right) 
=\left(\begin{array}{cc} a(k_{0}) & -e^{\sigma k_{0}}a(k_{0})\\
e^{\beta k_{0}}n(k_{0})/a(-k_{0})\hskip0.3cm & -e^{\sigma
k_{0}}n(k_{0})/a(-k_{0})\end{array} \right)\end{equation}
The expansion coefficients $C^{\ell}_{\alpha b}$ are defined in (2.15) as the $\ell$'th
derivatives of $V_{\alpha b}(k_{0})$ evaluated at $k_{0}=0$. The Bose-Einstein function in (A.1)
does not have finite derivatives at $k_{0}=0$. If  $a(k_{0})$ is chosen to have a pole at $k_{0}=0$
then the first row of (A.1) will have finite derivatives but not the second row. 
It is  easy  to evade this problem as follows.  As shown in Sec 6  it is only at the end of the
calculation, after calculating Green functions and taking the limit $N\to\infty$ that    $V_{\alpha
b}(k_{0})$ has to be (A.1). In the intermediate stages $V_{\alpha b}(k_{0})$ can differ from 
(A.1). One simple possibility is to introduce a fictitious chemical potential $\mu$ and set 
$n(k_{0})= 1/[\exp(\beta(k_{0}-\mu))-1]$. In intermediate stages of the calculations the
derivatives of $n(k_{0})$ will give inverse powers of $\mu$. 
At the end of the calculation,  the limit
$N\to\infty$ sums all the inverse powers of $\mu$ to give back  $1/[\exp(\beta(k_{0}-\mu))-1]$.
Thus one obtains a perturbative series containing  $1/[\exp(\beta(k_{0}-\mu))-1]$ in the
propagators. At that point it is safe to set $\mu\to 0$ to obtain the true theory. 
Note that since the particles under consideration are massive there are no infrared divergence.

\section{Residue Identities}

In the interaction picture the differential equation (5.6) for the field operators leads to the
frequency equation $0=\omega^{2}-E^{2}-h\omega^{2N}$. For h very small
there are two small  roots of the form $\omega\approx\pm( E-hE^{2N-1}/2)$. 
There are N-2 large roots $\omega\approx h^{-1/(2N-2)}\exp[i\pi\ell/(N-1)]$, where
$\ell=1,2,3,\dots, 2N-2$. The particular root with
$\ell=N-1$ is negative real; that with $\ell=2N-2$ is positive real.
 Thus of the total 2N roots,
there are two positive real, two  negative real, N-2 complex roots in the lower half-plane,
and N-2 complex roots in the upper half-plane. If the equation is modified to 
$0=\omega^{2}+i\eta-E^{2}-h(\omega^{2}+i\eta)^{N}$ for $\eta$ infinitessimal then the two
positive roots acquire small negative imaginary parts and the two negative roots acquire small
positive imaginary parts. Then there are N roots in the lower half-plane and N in the upper
half-plane. 

For $s=1,3,5,\dots,2N-3$ the following integral is convergent but  vanishes since the
integrand is odd:
\begin{eqnarray}
0=\int_{-\infty}^{\infty}
d\omega\;{\omega^{s}\over\omega^{2}\!+\!i\eta-E^{2}-h(\omega^{2}\!+\!i\eta)^{N}}.\nonumber
\end{eqnarray}
The contour can be closed in the lower half-plane. There is no contribution from the
 semicircular contour at infinity since $s\le 2N-3$. There are N roots $\omega_{j}$ in the lower
half-plane. By Cauchy's theorem and the definition (5.11) of the residues $Q_{j}$ the integral
gives \begin{equation}
0=\sum_{j=1}^{N}(\omega_{j})^{s}Q_{j}\hskip2cm (s=1,3,5,\dots,2N-3).\end{equation}
This result can be extended to higher values of $s$ as follows. Keep  $s$ odd but in the range 
$1\le s\le 2N-5$. Then multiply each $(\omega_{j})^{s}Q_{j}$  in (B.1) by 
\begin{equation}
E^{2}=(\omega_{j})^{2}-h(\omega_{j})^{2N}.\end{equation}
 The sum with $(\omega_{j})^{s+2}$
vanishes by the previous result. Hence the sum with $(\omega_{j})^{s+2N}$ also vanishes, where
$2N+1\le s+2N\le 4N-5$. No conclusion has been drawn about  the sum with $(\omega_{j})^{2N-1}$.
The same technique can be repeated. The complete result is
\begin{equation}
0=\sum_{j=1}^{N}(\omega_{j})^{s}Q_{j}\hskip2cm (s\; {\rm odd}, s\neq 2N\!-\!1, 4N\!-\!3,
 6N\!-\!5,\dots)
\end{equation}
To evaluate the sum at the first exceptional value, $s=2N-1$, 
use the identity
\begin{eqnarray}
{1\over\omega^{2}-E^{2}-h\omega^{2N}}=\sum_{j=1}^{N}Q_{j}
\,{2\omega_{j}\over\omega^{2}-\omega_{j}^{2}},\end{eqnarray}
which defines the residues $Q_{j}$. Set $\omega=0$ and multiply
both sides by $E^{2}$: \begin{eqnarray}
-1=-2\sum_{j=1}^{N}{Q_{j}\over \omega_{j}}E^{2}.\nonumber\end{eqnarray}
For each term in the sum use (B.2)
and the previous result (B.1) to get
\begin{eqnarray}
-1=2h\sum_{j=1}^{N}Q_{j}(\omega_{j})^{2N-1}\end{eqnarray}
To evaluate the sum at the next exceptional value of $s$,
multiply both sides of (B.4) by $E^{4}$,  apply $\partial/\partial\omega^{2}$, and
then set $\omega=0$:
\begin{eqnarray}
-1=-2\sum_{j=1}^{N}{Q_{j}\over (\omega_{j})^{3}}E^{4}.\nonumber\end{eqnarray}
For each term in the sum use (B.2) and the previous results (B.1) and (B.5) to get
\begin{equation}
-1=2h^{2}\sum_{j=1}^{N}Q_{j}(\omega_{j})^{4N-3}.\end{equation}
The next case is
\begin{equation}
-1=2h^{3}\sum_{j=1}^{N}Q_{j}(\omega_{j})^{6N-5}.\end{equation}

\end{document}